# Deep learning for magnitude prediction in earthquake early warning

Yanwei Wang[1], Zifa Wang[2,3], Zhenzhong Cao[1], and Jingyan Lan[1]

[1] Guangxi Key Laboratory of Geomechanics and Geotechnical Engineering, Guilin University of Technology, Guilin, China.

[2] College of Architecture and Civil Engineering, Henan University, Kaifeng, China

[3] Institute of Engineering Mechanics, China Earthquake Administration, Harbin, China.

Corresponding author: Z. F. Wang (zifa@iem.ac.cn)

**Key Points:**

- A self-adaptive deep learning approach is proposed for magnitude prediction.
- The approach is trained on a moderate-size data set and achieves better performance than other approaches.
- The research suggests that deep learning has great potential application in EEW.

**Abstract**

Fast and accurate magnitude prediction is the key to the success of earthquake early warning. We have proposed a new approach based on deep learning for P-wave magnitude prediction (EEWNet), which takes time series data as input instead of feature parameters. The architecture of EEWNet is adaptively adjusted according to the length of the input, thus eliminates the need of complicated tuning of hyperparameters for deep learning. Only the unfiltered accelerograms of vertical components are used. EEWNet is trained on a moderate number of data set (10,000s of records), but it achieves excellent results in magnitude prediction compared with approaches using parameters $\tau_{log}$, $\tau_c$ and $P_d$.

**Plain Language Summary**

After an earthquake occurs, earthquake early warning (EEW) sends warning information to service areas by using the first-arriving P-wave before the destructive S-wave arrives. Accurate earthquake magnitude prediction is the key challenge in EEW. In this paper, we applied the deep learning method in designing a neural network approach (EEWNet) to address those the challenge. After testing, EEWNet trained with a moderate number of data set (less than 20,000) has obvious advantages over traditional algorithms, with great application potential in future EEW systems.

## 1 Introduction

Most of magnitude prediction algorithms rely on relations between magnitude and manually defined parameters of initial P-wave(e.g. 3s), such as predominant period $\tau_p$ (Nakamura,1988), max predominant period $\tau_p^{max}$ (Allen & Kanamori, 2003), average period $\tau_c$ (Wu & Kanamori, 2005), the log-average period $\tau_{log}$ (Ziv, 2014), and peak displacement $P_d$ (Wu & Zhao, 2006). While many approaches have been attempted, the problem of magnitude prediction has not been well resolved probably due to the lack of whole fracture information and the limitations of the parameters themselves. Magnitude is a manually defined parameter related to the ground displacement. Using the initial P-wave as the input of deep learning to predict the maximum displacement of the complete record used to calculate magnitude seems to be an attractive proposal because of its direct applicability. The disadvantage of deep learning is that it usually requires a large labelled data set for training, which limits its application. In earthquake studies, there are millions of seismic records (seismograms and accelerograms) for small earthquakes ($M < 4$), while the records for medium to large earthquakes ($M >= 4$) applicable to EEW are limited in size, especially with added conditions defined by hypocentral distance, signal to noise ratio (SNR), peak amplitude and others. The challenge we face is, can deep learning approach trained by a moderate labelled data set be used to predict magnitude?

In this paper, we have developed a new deep learning approach (EEWNet) for magnitude prediction in EEW. EEWNet is built on convolution neural network, which automatically adjusts its architecture according to the length of input. The vertical accelerogram without any preprocessing is used as the input to EEWNet and the final maximum displacement is predicted. The magnitude can be calculated by the displacement. Tens of thousands of accelerograms labeled were used to train, validate and test EEWNet instead of a large data set in the order of millions that is often required by comparable approaches.

**2 Data and Methods**

2.1 Data sets

In this study, we used a high-quality data set that consists of 30,756 accelerograms recorded by 688 strong motion borehole sensors of the Kiban Kyoshin network (KiK-net) (10/1997~03/2019), with magnitude from 4 to 9 (3,648 earthquakes) and hypocentral distance between 25km and 200km. All accelerograms were vertical components and resampled to 100Hz. Two selection criteria were used to guarantee the high-quality of these accelerograms. One was the peak ground acceleration (PGA) over 2 gal, and the other was to check the data to avoid incomplete records, baseline drift records, and records containing multiple events (Figure S1). After that the signal to noise ratio (SNR) (Perol et al., 2018) of the data set is mostly greater than 10. We manually picked the P-wave arrival time to ensure accuracy. Each accelerogram was labeled for P-wave arrival time and the logarithm of the maximum displacement in the horizontal components. Selected accelerograms were divided into three data sets: 17,717 accelerograms from 10/1997 to 12/2011 were used as the training data set, 6,106 accelerograms from 01/2012 to 12/2014 were used as the validation data set, and 6,933 accelerograms from 01/2015 to 03/2019 were used as the test data set. The training and validation data sets were used for training and tuning hyperparameters of deep learning. The test data set was used to assess the performance of the deep learning after training.

2.2 EEWNet

The architecture of EEWNet is developed based on the one-dimensional convolution neural network, which is used to deal with the regression of a sequential data set. The architecture of EEWNet is composed of an input, multiple hidden layers, a fully connected layer and an output, as shown in Figure 1. For predicting magnitude, the input is the initial P-wave, and the output is the logarithm of the maximum displacement in the horizontal components of the complete records. EEWNet was trained by the training data set through optimization of the loss function defined as mean square error of output values. The self-adaptive architecture of the EEWNet were obtained by the performance of the training and validation data sets. For $2^N$ length of input, the total number of hidden layers is $N$, and each hidden layer is composed of a standard convolution operation consisting of $2^{L+3}$ filters and a pooling operation. $L$ is the serial number of hidden layer and equals to 1, 2, 3, …, $N$. For each filter, kernel size is 2, stride is 1 and padding type is 'same'. The activation function of convolutional layer uses rectified linear units (ReLUs) (Nair & Hinton, 2010). For each pooling optimization, method="max pooling", size=2 and stride=2. The keep probability of dropout for the last pooling layer is 0.5. The fully connected layer is a vector with $2^{N+3}$. The Adam stochastic optimization algorithm (Kingma & Ba, 2015) was used for the optimization with a learning rate of 0.0001. The batch size and epochs can be adjusted according to computer memory and computational convergence respectively. Here each batch consists of 200 accelerograms and epochs are 400. The architecture of EEWNet has four characteristics: the number of layer and filter size are adaptive to input; the length of input time series data can be less than $2^N$, but $2^N$ is recommended; each feature map in the last hidden layer is a value (1×1 dimension); and there are no normalization and regularization operations. EEWNet repeats convolution and pooling operations to achieve desired performance. For this study, EEWNet was programmed with TensorFlow GPU 1.6.

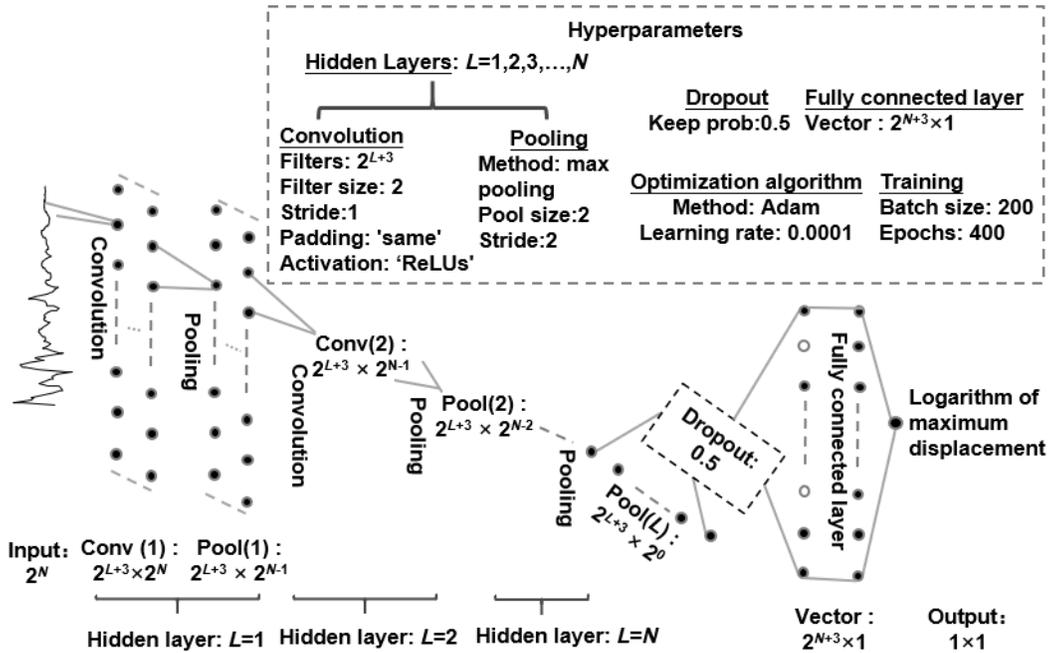

**Figure 1**. The architecture of EEWNet

## 3 Results

Magnitude of Kik-net given by Japan Meteorological Agency (JMA) is based on the maximum amplitudes of seismograms as shown into the following (Doi, 2014).

$$\log_{10}(A) = a \times M + b \times \log_{10}(R) + c \qquad (1)$$

where $A$ is maximum velocity or displacement of the complete record, $M$ is magnitude, $R$ is hypocentral distance, $a$, $b$ and $c$ are fitting coefficients. In EEW, magnitude is predicted by using formulas similar to equation (1), but it is based on the initial P-wave instead of the complete record, using parameters such as $\tau_c$, $\tau_{log}$ and $P_d$. $\tau_c$ and $\tau_{log}$ are frequency-based parameters that do not need to be corrected by hypocentral distance (Ziv, 2014). $P_d$ needs be corrected by hypocentral distance (Wu & Zhao, 2006). Here, initial P-wave of vertical accelerogram is used as the input of EEWNet to predict $\log_{10}(A)$. $A$ is the maximum displacement (μm, 1e-6m) in the horizontal components of the complete records (Doi, 2014). Magnitude can be then calculated by equation (1).

To evaluate the performance of EEWNet, $\tau_{log}$, $\tau_c$ and $P_d$, were also used to predict magnitude for comparison. The first 2.56s ($2^8$ samples) P-waves were used for the training, validation and test of EEWNet. The architecture of EEWNet was determined by $N=8$. The fitting coefficients $a$, $b$ and $c$ of $\tau_c$, $\tau_{log}$, $P_d$ and $\log_{10}(A)$ were estimated based on the training and validation data sets. The prediction errors of the test data set as a function of magnitude was shown in Figure 3. The standard deviation σ for $\tau_c$, $\tau_{log}$, $P_d$ and EEWNet was 0.82, 0.77, 0.68, and 0.40, respectively. The percentage of magnitude errors between -0.5 and 0.5 was 46.86%, 50.78%, 55.65% and 84.21% for $\tau_c$, $\tau_{log}$, $P_d$ and EEWNet, respectively, indicating EEWNet having much better magnitude prediction accuracy.

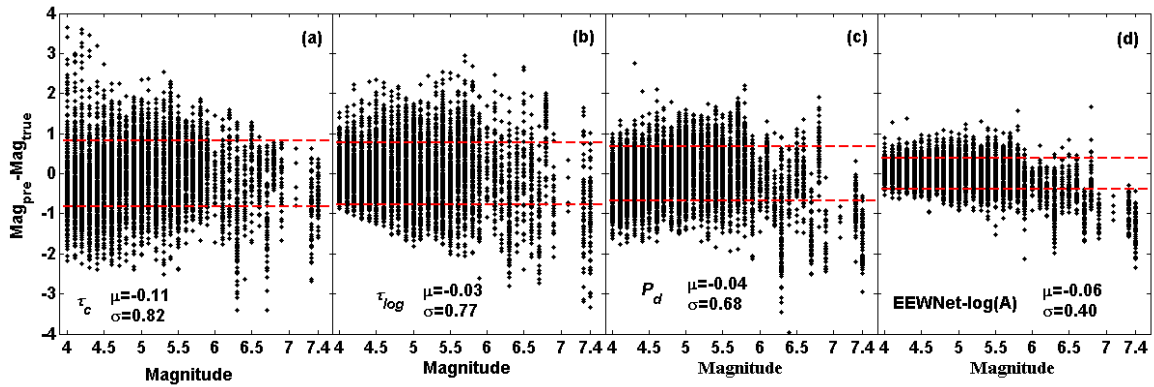

**Figure 3**. Magnitude prediction errors as a function of magnitude for $\tau_c$, $\tau_{log}$, $P_d$ and EEWNet. $\tau_{log}$, $\tau_c$ and $P_d$ were filtered using a 0.075Hz casual Butterworth high-pass filter with two poles. The calculation of $\tau_{log}$ was based on power spectrum of acceleration instead of velocity. $\mu$ and $\sigma$ were mean and standard deviation of errors. The black points indicated the errors of each prediction, and the red dashed lines were for $0 \pm \sigma$.

In EEW, magnitude needs to be continuously updated. To evaluate the performance of EEWNet with different lengths of initial P-wave, 64, 128, 512 and 1024 samples of initial P-wave were used to train and test respectively. The prediction errors of the test data set as a function of magnitude with different length of P-wave was shown in Figure 4. It is obvious that the standard deviation of prediction errors decreased with increasing input length, and the magnitude prediction of large earthquakes improved with the increased length of P-wave. To test for sample size other than $2^N$, 50, 100, 200 and 300 samples of initial P-wave were also used to train and test EEWNet, whose architecture was determined by inputs of length 64, 128, 256 and 512, respectively, and EEWNet achieved similar results.

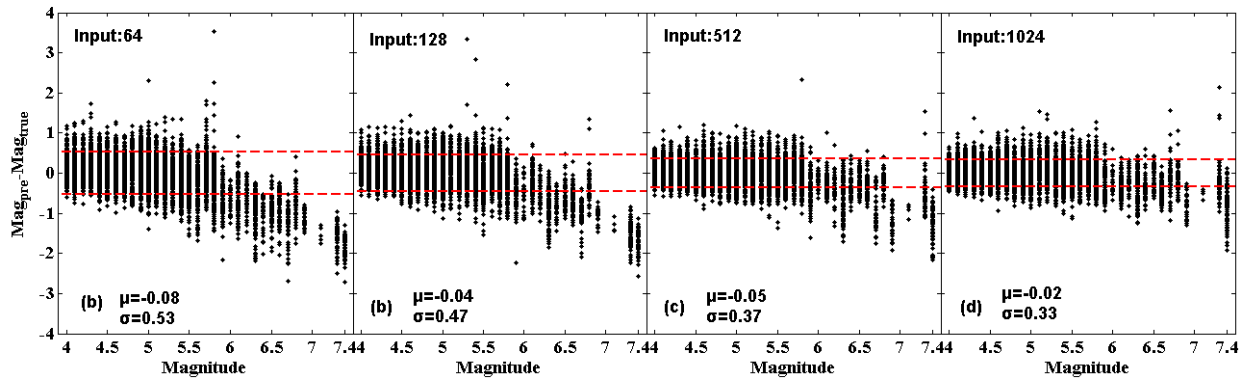

**Figure 4**. Magnitude prediction errors as a function of magnitude for EEWNet with different length of P-wave. The black points indicate the errors of each prediction, and the red dashed lines are the $0 \pm \sigma$.

## 4 Conclusions

We have proposed a new approach, EEWNet, based on convolutional neural network to predict magnitude for EEW. EEWNet used vertical accelerogram directly as an input without any preprocessing. The architecture of EEWNet is self-adaptive and it adjusts according to different length of inputs. This feature avoids hyperparameter tuning when the input length varies as in the case of magnitude prediction in EEW where P-wave length is increasing over time. For magnitude prediction, EEWNet was used to predict the maximum displacement in the horizontal components of the complete records to calculate the magnitude. The result was compared against those based on parameters $\tau_{log}$, $\tau_c$ and $P_d$ using the different lengths of initial P-wave. The comparison demonstrated that EEWNet achieved the highest precision with the smallest standard deviation. Moreover, the accuracy of magnitude prediction of EEWNet increased with increasing length of P-wave. The excellent performance of EEWNet shows its great application potential in EEW.

**Acknowledgments**

We would like to thank KiK-net online database for the recorded data (last accessed in March 2019). We are grateful to A. Ziv for sharing his source code of $\tau_{log}$. This research has been supported by the National Natural Science Foundation of China (Grant No. 51968016 and No. 51978634) and the Guangxi Innovation Driven Development Project (Science and Technology Major Project, Grant No. Guike AA18118008).